\documentclass[aps,twocolumn,pra,superscriptaddress,10pt]{revtex4-2}

\usepackage[colorlinks=true,urlcolor=blue,citecolor=blue,linkcolor=blue]{hyperref}
\usepackage{subfiles} 
\usepackage{enumerate}
\usepackage{xcolor}
\usepackage{dsfont}
\usepackage{amsfonts}
\usepackage{amsmath}
\usepackage{comment}
\usepackage{bm}
\usepackage{amssymb}
\usepackage{graphicx}
\usepackage[colorlinks=true,urlcolor=blue,citecolor=blue,linkcolor=blue]{hyperref}
\usepackage{tikz}
\usepackage{longtable}
\usepackage{tikz-cd}
\tikzcdset{
arrow style=tikz,
diagrams={>={Straight Barb[scale=0.8]}}
}

\DeclareGraphicsExtensions{.eps,.ps,.pdf}
\newcommand{\al}{\alpha}

\newcommand{\la}{\lambda}
\newcommand{\ka}{\kappa}

\newcommand{\La}{\Lambda}

\newcommand{\ket}[1]{\left| #1 \right\rangle}
\newcommand{\bra}[1]{\left\langle #1 \right|}

\newcommand{\ie}{\hbox{\em i.e.{}}}
\newcommand{\da}{\dagger}
\def \Tr {\text{Tr}}
\newcommand{\diff}{\mathop{}\!\mathrm{d}}
\newcommand{\rhs}{\hbox{r.h.s.{}}}

\newcommand{\Hs}{\mathcal{H}}
\newcommand{\Nega}{\mathcal{N}}

\usepackage{soul}
 
\begin{document}
\widetext

\title{Spin entanglement in antiferromagnetic spin-1 Bose-Einstein condensates}

\author{A. Santiago}
\email{g6\_sant17@ens.cnyn.unam.mx}
\affiliation{Departamento de F\'isica, Centro de Nanociencias y Nanotecnolog\'ia, Universidad Nacional Aut\'onoma de M\'exico\\
Apartado Postal 14, 22800, Ensenada, Baja California, M\'exico}

\author{F.{} Mireles}
\email{fmireles@ens.cnyn.unam.mx}
\affiliation{Departamento de F\'isica, Centro de Nanociencias y Nanotecnolog\'ia, Universidad Nacional Aut\'onoma de M\'exico\\
Apartado Postal 14, 22800, Ensenada, Baja California, M\'exico}

\author{E.{} Serrano-Ens\'astiga}
\email{ed.ensastiga@uliege.be}
\affiliation{Institut de Physique Nucléaire, Atomique et de Spectroscopie, CESAM, University of Liège
\\
B-4000 Liège, Belgium}
\begin{abstract}
We study the spin entanglement in a spin-1 Bose-Einstein condensate with antiferromagnetic atomic interactions using the Hartree-Fock approach. Based on the isomorphism between symmetric $N$-qubit states and spin-$j=N/2$ states, we analyze the negativity of the spin-1 ground state of the condensate viewed as a two spins $1/2$, 
and explore its dependence with the temperature and external Zeeman fields. The scope of this type of entanglement is highlighted and contrasted with other types of entanglement, as the mode and particle entanglement. It is shown that, at finite temperatures, there is a strong dependence of the negativity with respect to the strengths of quadratic Zeeman fields and the spin-spin interactions of the condensate. Interestingly, in the antiferromagnetic ground-state phase, the negativity and the linear Zeeman field are connected quadratically through the equation of a simple circle, in which its radius depends on the 
temperature. On the other hand, for the polar phase, being the phase that exhibits the highest degree of entanglement, we were able to identify a clear dependency on the spin-spin interactions and the Zeeman fields that can be expressed analytically in a closed form as a function of the temperature. This results might be relevant for applications in quantum information and metrology. 
\end{abstract}
\maketitle
\section{Introduction}
Spinor Bose-Einstein condensates (BECs) are many-body quantum systems of very diluted ultracold atomic gases that exhibit novel and exotic spin-dependent phases, making them an ideal platform for the study of complex spin interactions~\cite{lewenstein2012ultracold,Kaw.Ued:12,schafer2020tools,MISTAKIDIS20231}.  These systems are routinely confined in optical traps that allow their internal spin degree freedom to be manipulated with remarkable precision, leading to the realization of a variety of spin domain phases which are strongly influenced by the atomic species, besides the external fields involved~\cite{PhysRevLett.117.185302,PhysRevLett.119.050404,jimenez2019,prufer2020experimental}. Different ground states have been observed in different atomic species, such as the Polar phase in $^{23}$Na~\cite{stamper.Andrews.etal:1998,Jacob:12} condensates, or the Ferromagnetic phase in $^{87}$Rb~\cite{Chang.Hamley.etal:2004} and $^7$Li~\cite{PhysRev.2.033471} BECs. 

Spinor BECs are typically studied 
at utra-low temperatures, approaching absolute zero, including the recently temperature record of 37pK~\cite{PhysRevLett.127.100401}. At higher temperatures, and other factors such as interaction with the environment, turbulence, and quantum fluctuations, the BEC can be deviated from an ideal atomic condensate state~\cite{PhysRevLett.117.185302,PhysRevLett.119.050404,PhysRevA.108.053308}. For instance, as the temperature rises, the interactions of the cloud of thermally excited atoms with the condensate fraction of atoms become stronger~\cite{griffin2009bose,Kawa.Phuc.Blakie:2012,Ser.Mir:21}.
This leads to various novel physical phenomena such as magnetic spin domains, shifts in the phase boundaries, metastable phases and quantum quench dynamics~\cite{PRA.100.013622,PhysRevA.99.023606,PhysRevA.78.023632,PhysRevA.106.013304,PhysRevLett.126.063401,PhysRevLett.119.050404}. It is well-known that thermal effects can be also detrimental to certain many-particle quantum correlations, such as quantum entanglement. Hence, it is crucial to explore whether physical phenomena as quantum entanglement can still survive and be exploited at higher temperatures. 

Entanglement plays a paramount role in areas such as quantum metrology and quantum information~\cite{RevModPhys.90.035005,Nie.Chu:11,PhysRevA.108.010101}. 
Since a spinor BEC is essentially a many-body quantum system, it naturally conforms a suitable physics playground for the study of quantum entanglement. As a matter of fact, each particle in a BEC has an internal multimode degree of freedom, offering various forms of entanglement~\cite{PhysRevA.65.033619}. For instance, for an $N$-bosonic identical particle system describing a spin-1 condensate with hyperfine interactions, there can be an $N$-particle entanglement. However, since each particle conforms a three level system with internal levels, also called modes, given by $\ket{+1}$, $\ket{0}$, and $\ket{-1}$, can also form entangled states called mode entanglement~\cite{PhysRevA.65.033619}. This kind of entanglement is similar to the entanglement in optical modes of photonic systems.
Both types of entanglement, particle and mode entanglement are proposed to be useful for different applications~\cite{PhysRevA.67.013607,PhysRevLett.111.180401,mao2023quantum,PhysRevLett.125.033401}. 

Yet another type of entanglement is  the one that can be exhibited by a system of $N$ spin-$f$ states, and we call it spin-entanglement. As we will discuss in detail later, the Hilbert space of a spin-$f$ particle is isomorphic to the symmetric subspace of the Hilbert space associated to a system of $N=2f$ spin-1/2 states~\cite{PhysRevResearch.4.013178}. Therefore, we can reformulate quantities of multipartite quantum entanglement even in a single spin-$f$ atom through its corresponding symmetric $2f$ qubits of spin-1/2 states.
This correspondence will show to be practically useful for the description of spin entanglement in spinorial BECs with $f\geq 1$, including some practical applications in quantum metrology. As far we know, this idea of spin entanglement has not been discussed earlier. 

Our system of interest consists of a many-body system constituted by $N$ spin-1 atoms. The isomorphism between a spin-1 state as a symmetric bipartite spin-1/2 system allows the definition of spin entanglement. 
Whether this correspondence of spin-1 to two spins 1/2 is experimentally feasible or it remains just as an abstract equivalence, however, it is useful to study quantum correlations for applications in quantum technologies~\cite{PhysRevX.10.041012}. In fact, and as discussed further in the text, the measure of spin entanglement is linked to the Quantum Fisher Information with respect to an angular momentum operator, which is a relevant quantity in quantum-enhanced metrology of rotations~\cite{Bra.Cav:94,PhysRevA.82.012337,RevModPhys.90.035005}. 

Given these considerations, several questions naturally arise: To what extent do the many-particle interactions have influence on the spin entanglement of the atoms in a BEC? How do the thermal effects impact this entanglement? And, what are the optimal conditions for maximizing spin entanglement in a spinorial BEC? In this work, we address these questions going beyond mean-field theory by the use of the Hartree-Fock approach. We apply this formalism to explore the effects of particle interactions and the temperature on the spin entanglement in spin-1 BECs with antiferromagnetic interactions. Our theoretical results are valid for any general antiferromagnetic spin-1 BEC, and we apply them to study numerically a condensate gas of $^{23}$Na atoms.

The paper is organized as follows: In Sec.~\ref{Sec.Theory} we deal with the negativity concept, which is a measure of quantum entanglement for pure and mixed states. Then, we apply it to define the spin entanglement and comment on its potential applications. At the end of this section we briefly review the Mean-Field and the Hartree-Fock approximations, which are later employed in Sec.~\ref{Sec.Entanglement} to study the entanglement of the spinor BEC in its different spin phases. In Sec.~\ref{Sec.Ent.spin.spin}, we study the dependence of the spin entanglement with respect to the spin-spin interactions for the polar phase of the BEC. Lastly, we conclude in Sec.~\ref{Sec.Conclusions}.
\section{Theory}
\label{Sec.Theory}
\subsection{Spin entanglement and the Quantum Fisher Information} 
\label{Entanglement - Negativity}
Let us define first $\Hs_1$ as the Hilbert space of a single qubit system, here described by a spin-1/2 state, whose two-dimensional linear vector space is spanned by the basis vectors $\ket{\pm} =\ket{\frac{1}{2},\pm \frac{1}{2}}$, where $\ket{s,m}$ are the basis states  of the spin  operator $S_z$ with total spin $s=\frac{1}{2}$ and eigenvalues $m=\pm \frac{1}{2}$. Now, consider a two-qubit system $\Hs_1 \otimes \Hs_1$ and construct the so called computational basis defined by the tensor product of the single-qubit states $\{ \ket{++}, \ket{+-}, \ket{-+}, \ket{--} \}$, with $\ket{++} \equiv \ket{+}\otimes \ket{+}$ and analogously for the other basis elements. Another basis is formed by coupling the angular momentum of each qubit  such that there are two values of total angular momentum $l=0$ and $l=1$, $\Hs_1 \otimes \Hs_1 = \Hs^{(1)} \oplus \Hs^{(0)}$. $\Hs^{(l)}$ consists of the Hilbert space spanned by the states $\ket{l,m}$ corresponding to the eigenvectors of the total angular momentum operator $F_z$ with total angular momentum $l$ and magnetic quantum number $m$. Here $l=0, 1$ and $|m|\leqslant l$, with the symmetric and antisymmetric sectors corresponding to $l=1$ and $l=0$, respectively. The transformation between the computational basis $\{ \ket{++} , \ket{+-} , \ket{-+} , \ket{--} \}$ and the coupled basis $\{ \ket{1,1} , \ket{1,0} , \ket{0,0}, \ket{1,-1}  \}$ is given by the matrix
\begin{equation}
\begin{pmatrix}
1&0&0&0\\
0&\frac{1}{\sqrt{2}}&\frac{1}{\sqrt{2}} &0\\
0&\frac{1}{\sqrt{2}}&-\frac{1}{\sqrt{2}}&0\\
0&0&0&1\\
\end{pmatrix} .    
\end{equation}
An statistical mixture of quantum states, such as a quantum state at finite temperature, is described by a density matrix $\rho$, whose components in the computational basis can be denoted as $\rho_{m\nu\,n\mu} = \bra{m \mu} \rho \ket{n \nu}$ where the subindices $(m,n)$ and $(\mu ,\nu)$ label the single state in the first and second subsystem with $m,n,\mu,\nu = \pm$, respectively.

We now define the \emph{spin entanglement} in a two-qubit system using the entanglement measure of negativity between the qubit constituents~\cite{horodecki1996separability,PhysRevLett.77.1413,PhysRevA.65.032314,Bengtsson17}, which is expressed as follows:
\begin{equation}
\label{Neg.eq}
    \Nega (\rho) = \max \big[ 0, \, -2 \Lambda \big]  \, ,
\end{equation}
where $\Lambda$ is the minimum eigenvalue of the partial transpose of the state $\rho$, $\rho^{T_B}$, defined in the computational basis as $\rho^{T_{B}}_{m\mu,n\nu} = \rho_{m\nu ,n\mu}$.  

For spin-1 systems, its corresponding density matrix is restricted to the subspace with $l=1$, that is
\begin{equation}
    \rho^{\text{spin-1}} = \sum_{m_1 , m_2 =-1}^1 \rho_{m_1 m_2 } \ket{1,m_1} \bra{1,m_2} .
\end{equation}
This notion of spin entanglement differs from the \emph{mode entanglement} described in Ref.~\cite{PhysRevA.65.033619}. This can be seen even for a one-particle spin-1 state $\ket{10}$, which can be written in second quantization of three modes as
\begin{equation}
    \ket{10} = \ket{0,1,0}_{1,0,-1} = \ket{0}_1 \ket{1}_{0} \ket{0}_{-1},
\end{equation}
where $\ket{n}_{m}$ represents the state of $n$ particles in the mode $m$. The \rhs~of the last equation shows that $\ket{10}$ is a separable state, \ie, a state with mode entanglement equal to zero~\cite{PhysRevA.65.033619}. On the other hand, $\ket{10}$ in the two-qubit analogy is written in the computational basis as
\begin{equation}
\label{Eq.State.10}
\ket{10} = \frac{1}{\sqrt{2}} \Big( \ket{+-} + \ket{-+} \Big) ,
\end{equation}
that exhibits maximum spin entanglement $\Nega(\ket{10})= 1$.

Interestingly, the spin entanglement (negativity) of pure spin-1 states is proportional to other measures of entanglement. This connection arises due the uniqueness theorem of measures of entanglement for bipartite pure states~\cite{Bengtsson17}. To observe these equivalences, we can use the Schmidt decomposition of a two-qubit system $\Hs_1 \otimes \Hs_1 \equiv \Hs^A \otimes \Hs^B$, which reads
\begin{equation}
\label{Gen.Sta.Schmidt}
    \ket{\Psi} = \sum_{\alpha=1}^2 \sqrt{\Gamma_{\alpha}} \ket{\psi^A_{\alpha}}\ket{\psi^B_{\alpha}} ,
\end{equation}
where $\{ \ket{\psi^{a}_{1}} ,\ket{\psi^{a}_{2}} \}$ is a basis of $\Hs^a $ for $a=A,B$, and $\Gamma_{\alpha}$ are the Schmidt numbers with $\Gamma_1+ \Gamma_2=1$. The negativity of the previous state~\eqref{Gen.Sta.Schmidt} is simply reduced to~\cite{PhysRevA.65.032314}
\begin{equation}
    \Nega (\rho_{\Psi}) = \frac{1}{2}\left[ \left( \sum_{\alpha=1}^2 \sqrt{\Gamma}_{\alpha} \right)^2 -1 \right] = \sqrt{\Gamma_2 \left(1- \Gamma_2 \right)} , 
\end{equation}
where $\rho_{\Psi}=\ket{\Psi}\bra{\Psi}$. Another measure is the linear entropy of entanglement $E_L$, which in terms of the negativity is given by~\cite{Bengtsson17}
\begin{equation}
    E_L(\rho_{\Psi}) = 2\left( 1- \sum_{\alpha}^2 \Gamma_{\alpha}^2 \right) = 4\Gamma_2 \left( 1- \Gamma_2 \right) = 4\Nega^2 .
\end{equation}
Note that for a two-qubit system, $E_L$ coincides with the so-called tangle. Besides, it is related to the measure of concurrence of the state through $C \equiv \sqrt{E_L} = 2 \Nega$~\cite{Bengtsson17}.
All the previous results remain valid for symmetric two-qubit states (\ie, spin-1 state). In addition, it is fulfilled that $\ket{\psi_{\alpha}^A}= \ket{\psi_{\alpha}^B} \equiv \ket{\psi_{\alpha}}$ due to its permutation invariance. 

There is yet another connection between the negativity and a well known physical quantity, namely the Quantum Fisher Information (QFI). To see this, first we use the Bloch representation of the spin-1/2 states that identifies any state with a point on the surface of a unit sphere~\cite{Nie.Chu:11,Bengtsson17}. In particular, we can write the first state $\ket{\psi_1}$ of the expansion~\eqref{Gen.Sta.Schmidt} as
\begin{equation}
\ket{\psi_1} = \ket{\mathbf{n}} = \cos \left( \frac{\theta}{2} \right) \left| + \right\rangle + e^{i \phi} \sin \left( \frac{\theta}{2} \right) \left| -\right\rangle ,    
\end{equation}
where $(\theta, \phi)$ are the spherical angles of the unitary vector $\mathbf{n}$. Since, $\ket{\psi_2} \perp \ket{\psi_1}$, we obtain that
\begin{equation}
\ket{\psi_2} = \ket{-\mathbf{n}} = e^{i \delta} \left[ \sin \left( \frac{\theta}{2} \right) \left| + \right\rangle - e^{i \phi} \cos \left( \frac{\theta}{2} \right) \left| -\right\rangle  
\right] .
\end{equation}
These states satisfy the equation $(\mathbf{n} \cdot \mathbf{F}^{(1/2)}) \ket{\pm \mathbf{n}} = \pm \frac{1}{2} \ket{\pm \mathbf{n}}$, where $\mathbf{F}^{(f)}$ represents the angular momentum operators for spin-$f$ states. Hence, a symmetric two-qubit state~\eqref{Gen.Sta.Schmidt} can be rewritten as
\begin{equation}
\label{PsiS}
    \ket{\Psi} = \sqrt{\Gamma_1} \ket{\mathbf{n}} \ket{\mathbf{n}} + \sqrt{\Gamma_2} \ket{-\mathbf{n}} \ket{-\mathbf{n}} .
\end{equation}
From this, we can conclude that the maximally entangled states, characterized by $\Gamma_1 = \Gamma_2 = 1/2$, are analogous to NOON or GHZ (Greenberger–Horne–Zeilinger) states, which possess valuable properties for quantum information and quantum metrology~\cite{PhysRevLett.91.107903,RevModPhys.90.035005,PhysRevA.82.012337}. These states are also known to maximize the QFI parameter for infinitesimal rotations in a given direction. Specifically, the QFI parameter of a state $\ket{\Psi}$, with respect to the angular momentum operator $\mathbf{n} \cdot \mathbf{F}$ and denoted as  $I_{\mathbf{n}}$, is proportional to the variance of  $\mathbf{n} \cdot \mathbf{F}$
\begin{equation}
\begin{aligned}    
    I_{\mathbf{n}} (\ket{\Psi}) & \equiv 4 \left[ \Big\langle \left( \mathbf{n} \cdot \mathbf{F}^{(1)} \right)^2 \Big\rangle - 
    \Big\langle \mathbf{n} \cdot \mathbf{F}^{(1)} \Big\rangle^2 \right] 
\\
& =4\left[  1 - \left( \Gamma_1 - \Gamma_2 \right)^2 \right]
\\
& =16 \Nega^2 .
\end{aligned}
\end{equation}
Consequently, $I_{\mathbf{n}}$ is maximized for the case discussed in \eqref{PsiS}  since $\Gamma_1 = \Gamma_2 = 1/2$. Now, for a number $M$ of measurements, and according to the Quantum Cramer-Rao bound~\cite{Bra.Cav:94,helstrom1969quantum}, the estimation of an infinitesimal rotation $\eta$ along the $\mathbf{n}$ axis for a state $\ket{\Psi}$, denoted as $\text{Var} \left[ \hat{\eta} \right]$ where $\hat{\eta}$ is the estimator of $\eta$, is inversely proportional to the 
QFI, that is
\begin{equation}
\text{Var} \left[ \hat{\eta} \right] \geqslant \frac{1}{M I_{\mathbf{n}}(\ket{\Psi})}.
\end{equation}
Therefore, any spin-1 pure state with maximum spin entanglement will hold to be an optimal state to estimate an infinitesimal angle in a specific direction. 

Let us now present some some examples to illustrate our previous discussion. First, consider the state $\ket{\Psi}= \ket{10}$ (See Eq.~\eqref{Eq.State.10}) and notice that it can also be expressed as
\begin{equation}
\label{Eq.Sta.10}
    \ket{10} =\frac{1}{\sqrt{2}} \Big( \ket{\mathbf{x}}\ket{\mathbf{x}} - \ket{-\mathbf{x}}\ket{-\mathbf{x}}  \Big) .
\end{equation}
Clearly, $\ket{\Psi}$ maximizes $I_{\mathbf{x}}$ and, consequently, serves as the optimal state for the estimation of a rotation about the $\mathbf{x}$ axis, despite having mode entanglement null~\cite{PhysRevA.65.033619}. Another example is the state
\begin{equation}
\label{Eq.Sta.Super}
   \ket{\Psi'} = \frac{1}{\sqrt{2}} \Big( \ket{11} + \ket{1-1} \Big)=\frac{1}{\sqrt{2}} \Big( \ket{\mathbf{z}}\ket{\mathbf{z}} + \ket{-\mathbf{z}}\ket{-\mathbf{z}}  \Big), 
\end{equation}
which is the most susceptible state to detect a rotation about the $\mathbf{z}$ axis. In contrast, this state $\ket{\Psi'}$ maximizes both mode and spin entanglement types. These two examples demonstrate that spin entanglement effectively identifies all states that are optimal for rotation detection. From now on, we only work with the spin entanglement defined by the negativity, referring to it simply as entanglement. The advantage of the negativity is that it is a well-defined measure of entanglement for pure and mixed states~\cite{PhysRevA.65.032314,Bengtsson17}. 
\subsection{Spinor Bose-Einstein Condensates}
The system under consideration is a spinor BEC conformed by $N=10^4$ atoms of $^{23}$Na, each with total angular momentum (spin) 1. The full Hamiltonian in the second-quantization formalism describing the atomic gas subject to linear ($p$) and quadratic ($q$) Zeeman fields along the $z$ axis reads
\begin{align}
\label{Full.Ham}
& 
\nonumber
\hat{H} 
= \int \diff \mathbf{r}  \Bigg(  \hat{\bm{\Psi}}^{\da} 
\left( h_{s} \mathds{1}_3
 - p F_z + q F_z^2 
\right) \hat{\bm{\Psi}} 
\\
& 
+ \frac{c_0}{2} \sum_{i,j}
 \hat{\psi}_i^{\da} \hat{\psi}_j^{\da} \hat{\psi}_j \hat{\psi}_i
 + \frac{c_1}{2} \sum_{\al,i,j,k,l} 
(F_{\al})_{ij} (F_{\al})_{kl} \hat{\psi}_i^{\da} \hat{\psi}_k^{\da} \hat{\psi}_l \hat{\psi}_j
\Bigg) , 
\end{align}
with $(c_0 , c_1) =$ (0.42$\,a_{B}^3$eV,\,$c_0/27)$ for the case of $^{23}$Na atoms, where $a_{B}$ is the Bohr radius~\cite{Ser.Mir:21}. The $3\times3$ matrices $F_{\alpha}$ represent angular momentum operators for spin-1 scaled by $\hbar$, where $\alpha = x,y,z$. Consequently, the $F_{\al}$ matrices are dimensionless. Atomic gases with $c_1>0$, such as $^{23}$Na atoms, are said to have antiferromagnetic or polar spin-spin interactions~\cite{Kaw.Ued:12}. 
\subsubsection{Zero temperature case: Mean-Field approximation}
\begin{figure}[t!]
    \centering    
    \includegraphics[width=0.45\textwidth]{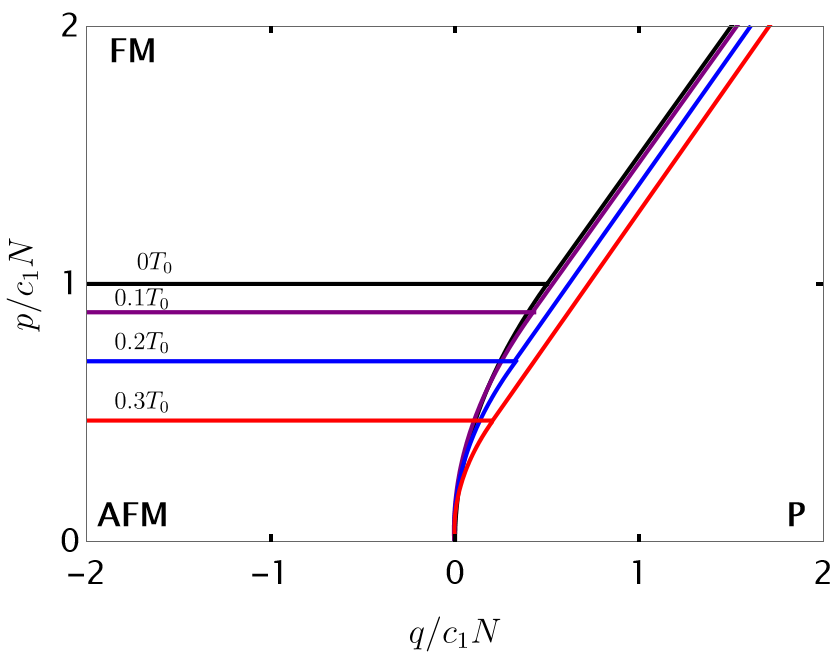}
    \caption{Phase diagram of a spin-1 BEC of $^{23}$Na atoms at $T/T_0=0, 0.1,0.2$ and 0.3.}
    \label{Fig.1}
\end{figure}

Mean-field (MF) approximation assumes the replacement of  $\hat{\bm{\Psi}} \rightarrow \langle \hat{\bm{\Psi}} \rangle = \bm{\Phi}$ , where $\bm{\Phi} = (\phi_f , \phi_{f-1} , \dots , \phi_{-f} )^{\text{T}} $ is the spinor order-parameter obeying $\bm{\Phi}^{\dagger} \bm{\Phi} = N$~\cite{Kaw.Ued:12,lewenstein2012ultracold}. The spin phase of the BEC is found by minimizing the energy functional $E[\bm{\Phi}]= \langle \hat{H} \rangle$. For a BEC of $^{23}$Na atoms, the phase diagram in the external parameters $(q,p)$ has the following (Fig.~\ref{Fig.1}) spin phases (See Refs.~\cite{Kawa.Phuc.Blakie:2012,Ser.Mir:21} for more details): 

(1) Ferromagnetic (FM) phase: Its order parameter is $\bm{\Phi}_{FM} = \sqrt{N} \ket{1,1} =  \sqrt{N} \ket{++}$. Since the state is a tensor product of single states, the FM phase do not possess entanglement $\Nega ( \rho_{FM} )=0$ with $\rho_{FM} = \bm{\Phi}_{FM} \bm{\Phi}^{\dagger}_{FM} $. The state $\ket{1,-1} = \ket{--}$ is also a FM phase but oriented in the $-z$ direction. However, it only appears as a spin phase for $p\leqslant 0$.

(2) Polar (P) phase: Here $\bm{\Phi}_{P} = \sqrt{N} \ket{1,0} $. Opposed as the FM phase, $\bm{\Phi}_{P}$ is maximally entangled with $\Nega ( \rho_P ) = 1$. This state is also equal to Eq.~\eqref{Eq.Sta.10} that maximizes $I_{\mathbf{x}}$.

(3) Antiferromagnetic (AFM) phase: It consists of a family of quantum states with order parameter depending on the parameter $p$
\begin{equation}
\label{AFMOrdParam}
    \bm{\Phi}_{AFM} = \sqrt{\frac{N(1+\bar{p})}{2}} \ket{1,1} + \sqrt{\frac{N(1-\bar{p})}{2}} \ket{1,-1} .
\end{equation} 
with $\bar{p}= p/ c_1 N$. The phase is well defined only for $\bar{p} \in [0 , 1]$ and its negativity is
\begin{equation}
\label{Eq.Nega.AFM.T0}
\Nega( \rho_{AFM})= N\sqrt{1- \bar{p}^2}.    
\end{equation}
In particular, $\bm{\Phi}_{AFM}$ is equal to the P phase but oriented in the $y$ axis at $\bar{p}=0$ and equal to the FM phase at $\bar{p}=1$. Additionally, the AFM phase for $\bar{p}=0$ is equal to the state~\eqref{Eq.Sta.Super} that maximizes $I_{\mathbf{z}}$.

We remark that the entanglement is independent of the parameters $q$ and $c_0$ for all spin phases, while it depends on $c_1$ exclusively in the AFM phase. 
\subsection{Hartree-Fock method}
\label{SubSec.HF}
The Hartree-Fock (HF) approximation~\cite{blaizot1986quantum,griffin2009bose,Kawa.Phuc.Blakie:2012}, despite of being the simplest many-body theory after the mean-field approximation, allows us to study the BEC at finite temperatures. Formally, the field operator is given by the order parameter, used in the MF approximation, and a perturbation $\hat{\delta}_j$, $\hat{\psi}_j = \phi_j + \hat{\delta}_j$. For simplicity, we neglect the three-field correlations $\langle  \hat{\delta}_i \hat{\delta}_j \hat{\delta}_k^{\da} \rangle$ (Hartree-Fock-Bogoliubov approximation) and the anomalous density $\langle \hat{\delta}_i \hat{\delta}_j \rangle$ (Popov approximation) \cite{griffin2009bose}, which gives a reasonable first approximation for diluted gases at all temperatures below the critical temperature \cite{Griffin:96,Pro2008}. We now apply the elements of the HF theory following Ref.~\cite{Ser.Mir:21}. 

The condensate (\emph{c}) and non-condensate (\emph{nc}) atoms are represented by their respective density matrices $\rho^c_{ij} =N^c \phi^*_j \phi_i $ and $ \rho^{nc}_{ij}= \langle \hat{\delta}_j^{\da} \hat{\delta}_i \rangle $. The trace of each density matrix corresponds to the number of atoms of each fraction, $\Tr (\rho^a) = N^a$ for $a=n, nc$. The non-condensate atoms $\rho^{nc}$ behave as a cloud of thermally excited atoms that interacts non-trivially with the condensate fraction $\rho^c$. The total system is then represented by $\rho =\rho^{c}+\rho^{nc}$ with $\Tr \rho = N = N^c +N^{nc}$. 
The HF energy of $\rho$, which incorporates a Lagrange multiplier to conserve its total number of particles $\mu ( N -\Tr \rho)$, has the form
 \begin{align}
\label{HF.energy}
E_{HF} = &  E_{s} + \Tr \left[ \rho \left( -p F_z + q F_z^2 \right) \right] -\mu \left( \Tr \rho- N \right)
\nonumber
\\
&
+ \frac{c_0}{2}\Big( N^2 + \Tr \left[ \rho^{nc} \left(2\rho^c + \rho^{nc} \right) \right] \Big)
\\
\nonumber
&
+ \frac{c_1}{2} \sum_{\alpha} \Big(  \Tr \left[ \rho F_{\al} \right]^2 
+ \Tr \left[ F_{\al} \rho^{nc} F_{\al} \left( 2 \rho^c + \rho^{nc} \right) \right] \Big) ,
\end{align}
where the trace involves a summation over both spatial and spinor quantum numbers. For $U(\mathbf{r})=0$, the spatial quantum number is the wavevector $\mathbf{k}$, and the spatial energy consists of only the kinetic energy represented by $E_{s}$. On the other hand, the two-body interactions have two terms, the direct and exchange interactions. 

The condensate fraction of the system $\rho^c=N^c \bm{\Phi} \bm{\Phi}^{\da}$ is a pure state with $\mathbf{k}=\mathbf{0}$. Hence, the resulting GP equations $\delta E_{HF}/\delta \phi^*_m =0 $ are given by a system of three (non-linear) equations involving $\phi_m$ and $\rho^{nc}$. On the other hand, $\rho^{nc}$ is written as a sum of its eigenvectors $\bm{\xi}^{\la} = (\xi_1^{\la}, \xi_0^{\la}, \xi_{-1}^{\la})^{\mathrm{T}}$ weighted by their Bose-Einstein distribution factor $n_{\la}$, \begin{equation}
\rho^{nc}_{ij} = \sum_{\la} n_{\la} \xi^{\la}_i \xi^{\la*}_j  
\, , \quad
n_{\lambda} =\left( e^{\beta \epsilon_{\la}} -1 \right)^{-1}
\, .
\end{equation}
 The global subindex $\la$ includes the spatial and spinor quantum numbers, $\la= (\mathbf{k}, \nu)$, with $\nu=1,2,3$ and $\beta= 1/k_{B}T$ where $k_{B}$ is the Bolztmann constant and $T$ the absolute temperature. The eigenvectors $\bm{\xi}^{\la} $ and their associated energies $\epsilon_{\la}$ are obtained by the non-condensate Hamiltonian $A$, given by $A_{ij} = \delta E_{HF}/\delta \rho^{nc}_{ji} $. The decoupling of the spatial and spinor parts in the Hamiltonian $A$ leads to $\epsilon_{\lambda}= -\hbar^2 k^2/2M + \kappa_{\nu}$, with $M$ the $^{23}$Na atomic mass and $\kappa_{\nu}$ the eigenvalue of the spinor part of $A$. The spatial part of $\rho^{nc}$ can be integrated using that $\sum_{\mathbf{k}}\rightarrow (2\pi)^{-3} \int\diff \mathbf{k}$,
\begin{equation}
\label{Poly.sta}
\rho^{nc}_{ij} = \sum_{\nu=1}^3 \xi^{\nu}_i \xi^{\nu *}_j  \frac{Li_{3/2}\left(e^{-\beta \kappa_{\nu}} \right)}{\la_{dB}^3}  \, ,
\end{equation}
where $Li_{3/2}(z)$ is the polylogarithm function and $\la_{dB} = h / \sqrt{2\pi M k_B T} $ is the thermal de Broglie wavelength. The eigendecomposition of $A_{ij}$, which is now a $3\times 3$ matrix, are called the HF equations.

To simplify the derivations and approximations, we scale to the dimensionless variables per particle
\begin{equation*}
\left( 
\begin{array}{c}
\bar{p} \\ \bar{q} \\ \bar{L} \\ \bar{\mu} \\ \bar{A} \\ \bar{\kappa}_{\nu}
\end{array}
 \right) = \left( 
\begin{array}{c}
p \\ q \\ L \\ \mu \\ A \\ \kappa_{\nu}
\end{array}
 \right) / c_1 N \, , \quad
\left( 
\begin{array}{c}
\bar{N}^{c} \\ \bar{N}^{nc} 
\\ \bar{\rho}^{c} \\ \bar{\rho}^{nc} \\ \bar{\rho} 
\\ \bar{\Nega}
\end{array}
 \right) = \left( 
\begin{array}{c}
N^c \\ N^{nc} \\ \rho^{c} \\ \rho^{nc} \\ \rho \\ \Nega
\end{array}
 \right) / N  .
\end{equation*}
We additionally define
\begin{equation}
\label{Sca.Eqs}
\bar{\bm{\Phi}} = \frac{\bm{\Phi}}{\sqrt{N}} ,
\quad
\bar{c}_0 = \frac{c_0}{c_1} , 
 \quad 
\bar{T} = \frac{T}{T_0} ,
\end{equation}
with $T_0$ being the critical temperature for an scalar BEC, $T_0=3.31 \hbar^2 N^{2/3}/k_B M$~\cite{pethick2008bose}. The condensate and non-condensate fractions satisfy that $\bar{N}^c+\bar{N}^{nc}=1$.  $\bar{\rho}^{nc}$ is written as 
\begin{equation}
\label{rhonc.bar}
\bar{\rho}^{nc}_{ij} = \sum_{\nu} \Lambda_{\nu} \xi^{\nu}_i \xi^{\nu*}_j 
\, , \quad \Lambda_{\nu} = \frac{Li_{3/2}\left(e^{-z_{\nu}} \right)}{N\la_{dB}^3} ,
\end{equation}
with $\bar{N}^{nc}= \sum_{\nu} \Lambda_{\nu}$ and
\begin{equation}
z_{\nu}=  \left( \frac{c_1 N}{k_B T_0} \right) \left( \frac{\bar{\kappa}_{\nu}}{\bar{T}} \right) = k_1 \frac{\bar{\kappa}_{\nu}}{\bar{T}}  .
\end{equation}
Unless it is stated otherwise, we will work with the scaled variables but suppressing the bar symbol in each term.
We write the resulting GP-HF equations
\begin{equation*}
\mu \bm{\Phi} = L \bm{\Phi}  ,
\end{equation*}
\begin{equation}
\begin{aligned}
\label{sca.GPE.eq}
 L = 
& -p F_z + q F_z^2+ c_0 \big( \mathds{1}_3  + \rho^{nc} \big)
 \\
& + \sum_{\al} \Big\{ \Tr \left[F_{\al} \rho \right] F_{\al} + F_{\al} \rho^{nc} F_{\al} \Big\} \, ,
 \\
 A= & -\mu \mathds{1}_3 - p F_z + q F_z^2 + c_0 \big( \mathds{1}_3 + \rho \big)
\\
&
 + \sum_{\al} \Big\{
 \Tr \left[ \rho F_{\al} \right] F_{\al} + F_{\al} \rho F_{\al} \Big\} \, .
\end{aligned}
\end{equation}
Usually, the GP-HF equations \eqref{sca.GPE.eq} are solved self-consistently \cite{blaizot1986quantum,griffin2009bose,Kawa.Phuc.Blakie:2012}. However, we can obtain analytic approximations following the method used in Ref.~\cite{Ser.Mir:21}, and its implementation for general spinor BEC in Ref.~\cite{Ser.Mir2:23}. The method exploits the rotational symmetries of the phases to reduce the degrees of freedom. For completeness, we review briefly the approximations and expressions of the density matrices. The first approximation is given to $ \Lambda_{\nu} = Li_{3/2}\left(e^{-z_{\nu}} \right)/N\la_{dB}^3 $
\begin{equation}
\label{approx.Li}
\begin{aligned}    
 \Lambda_{\nu} 
\approx &
\frac{1}{N\la_{dB}^3} \Bigg[ \zeta\left( \frac{3}{2} \right) - 2\sqrt{\pi z_{\nu}} 
- \zeta\left( \frac{1}{2} \right) z_{\nu} + \mathcal{O}\left( z_{\nu}^{2}\right) \Bigg]  .
\end{aligned}
\end{equation}
The latter approximation is valid for our case because we are interested in the qualitative behavior around $ k_1 \kappa_{\nu} / T \approx 0 $ where $k_1 = c_1N/k_B T_0=1.75 (10^{-3})$. Here $\zeta(z)$ is the Riemann zeta function. For simplicity, we write the expansion as
\begin{equation}
\label{LambdaGenExp}
    \La_{\nu} = 
    k_2 \left[
\zeta \left( \frac{3}{2} \right) T^{3/2} 
-2 T \sqrt{\pi k_1  \ka_{\nu}} 
- \zeta \left( \frac{1}{2} \right) k_1  \ka_{\nu} \sqrt{T}
\right] ,
\end{equation}
with $k_2 =1/ \la_{0}^3 N \approx 0.382358$ and where $\la_0$ is the thermal de Broglie wavelength at $T=T_0$. The last expression gives a non-transcendental expression between the fraction $\Lambda_{\nu}$ and the HF energy $\kappa_{\nu}$. We can write now the GP-HF equations which are considerably reduced for each phase. In particular, the eigenenergies and the eigenvectors of $\rho^{nc}$ for the FM, P and AFM phases can be written analytically in terms of the physical parameters after some feasible approximations. We take the following equations from Ref.~\cite{Ser.Mir:21} where they were used to calculate the regions of metastability of each phase.
%
%
%
%
%
%
%
\subsubsection{FM phase}
The eigenvectors of $\rho^{nc}$ are given by $\ket{1,m}$, and $\rho^c$ is defined by $\ket{1,1}$. Hence,
\begin{equation}
\label{Density.matrix}
  \rho = \sum_{m=-1}^1 f_m \ket{1, \, m} \bra{1, \, m} \, ,
\end{equation}
with $f_1 = N^c + \La_1 = 1- \La_0 - \La_{-1}$, $f_0 = \La_0$ and $f_{-1} = \La_{-1}$. The eigenenergies of the non-condensate fraction are (see Ref.~\cite{Ser.Mir:21} for more details),
\begin{align}
\kappa_1^{(FM)} = & \left(c_0 + 1 \right) \left( 1- \Lambda_1 - \Lambda_0 - \Lambda_{-1} \right)  \, ,
\nonumber
\\
\kappa_0^{(FM)} = & p-q -\left( c_0 + 1 \right) \Lambda_1 + \left( c_0 - 1 \right) \Lambda_0 + 2 \Lambda_{-1} ,
\label{kappa.FM}
\\
\kappa_{-1}^{(FM)} = & 2(p-1) -\left( c_0 + 1 \right) \Lambda_1 + 2 \Lambda_0 + \left( c_0 + 5 \right) \Lambda_{-1}  .
\nonumber
\end{align}
Analytical expressions can be obtained for $\ka_{m}^{(FM)}$ up to $\mathcal{O}(k_1^{2}, T^{3/2})$~\cite{Ser.Mir:21}
\begin{equation*}
\kappa_1^{(FM)} 
= c_0 +1 + F_{-}(T) ,
\quad
\kappa_0^{(FM)}= 
p-q+ F_{+}(T) ,
\end{equation*}
\begin{equation}
\label{FMkappa.aprox}
\kappa_{-1}^{(FM)}= 2\left(p- 1 \right) + F_{+}(T)
,
\end{equation}
with
\begin{equation}
    F_{\pm}(T)= 
k_1 k_2 (c_0+1 )^2 \zeta \left(\frac{1}{2}\right)  \sqrt{T}
\pm 2 \sqrt{\pi k_1 (c_0+1 )^3 }  k_2 T 
. 
\end{equation}
For this case and in our approximations, $f_0 , f_{-1} \ll 1$. Specifically, the greatest values of $f_0$ and $f_{-1}$ are about 0.1 for the $T=0.3$~\cite{Ser.Mir:21}.
\subsubsection{P phase}
The density matrix $\rho$ of the P phase has the same eigenvectors as the FM phase. Consequently, $\rho$ is equal to Eq.~\eqref{Density.matrix} but with 
$f_0 = N^c + \La_0 = 1- \La_1 -\La_{-1}$ and $f_{\pm} = \La_{\pm}$. The HF energies $\ka_{m}^{(P)}$ satisfy that
\begin{align}
\kappa_1^{(P)} = & q-p+1 +c_0 \left( \Lambda_1 - \Lambda_0 \right) - 3 \Lambda_{-1} ,
\nonumber
\\
\kappa_0^{(P)} = & c_0 \left(1- \Lambda_1 - \Lambda_0 - \Lambda_{-1} \right)  ,
\label{kappa.P}
\\
\kappa_{-1}^{(P)} = &q+p+1 -3 \Lambda_1 + c_0 \left(-\Lambda_0 + \Lambda_{-1} \right)  .
\nonumber
\end{align}
Up to $\mathcal{O}(k_1^2 , T^{3/2})$, $\kappa_{m}^{(P)}$ are expressed by~\cite{Ser.Mir:21}
\begin{align}
\kappa_{\pm1}^{(P)} 
= \mp p + q + 1 + G(T) ,
\quad 
\kappa_0^{(P)} 
= c_0 + G(T) ,
\label{Pkappa.aprox}
\end{align}
with
\begin{equation}
    G(T)= c_0^2 k_1 k_2 \zeta \left(\frac{1}{2}\right)  \sqrt{T}
+2 \sqrt{\pi k_1 c_0^3} k_2 T .
\end{equation}
Here, $f_{\pm 1} \ll 1$~\cite{Ser.Mir:21}, with maximum values of $f_{\pm 1} \approx 0.1$ for $T=0.2$.
\subsubsection{AFM phase}
The family of states of the AFM phase are given by the order parameter, $\bm{\Phi}= (\phi_1 , \phi_0, \phi_{-1})=(\cos \chi, 0, \sin \chi)^{\text{T}}$ with $\chi \in (0,\pi/4 ]$. Hence, the BEC gas $\rho$ and its fractions $\rho^c$ and $\rho^{nc}$ must possess the same symmetries, implying that
\begin{equation}
\begin{aligned}
\label{AfmDensMat}
\rho = & \rho^c + \rho^{nc} \, , 
\\ 
\rho^{c} = & N^c \left(
\begin{array}{c c c}
\cos^2 \chi & 0 & \cos \chi \sin \chi \\
0 & 0 & 0 \\
\cos \chi \sin \chi & 0 & \sin^2 \chi 
\end{array}
\right) \, ,
\\
\rho^{nc} = & \left(
\begin{array}{c c c}
a & 0 & D \\
0 & b & 0 \\
D & 0 & c
\end{array}
\right) \, ,
\end{aligned}
\end{equation}
where $D$ is a real number. Although there are approximate expressions for the aforementioned variables (see Ref.~\cite{Ser.Mir:21}), they are too lengthy and not too illuminating. We therefore restrict ourselves to the study of AFM phase numerically. 
\section{Entanglement in spin-1 BEC at finite temperatures}
\label{Sec.Entanglement}
%
%
\begin{figure}[t!]
    \centering        \includegraphics[width=0.5\textwidth,height=\textheight,keepaspectratio]{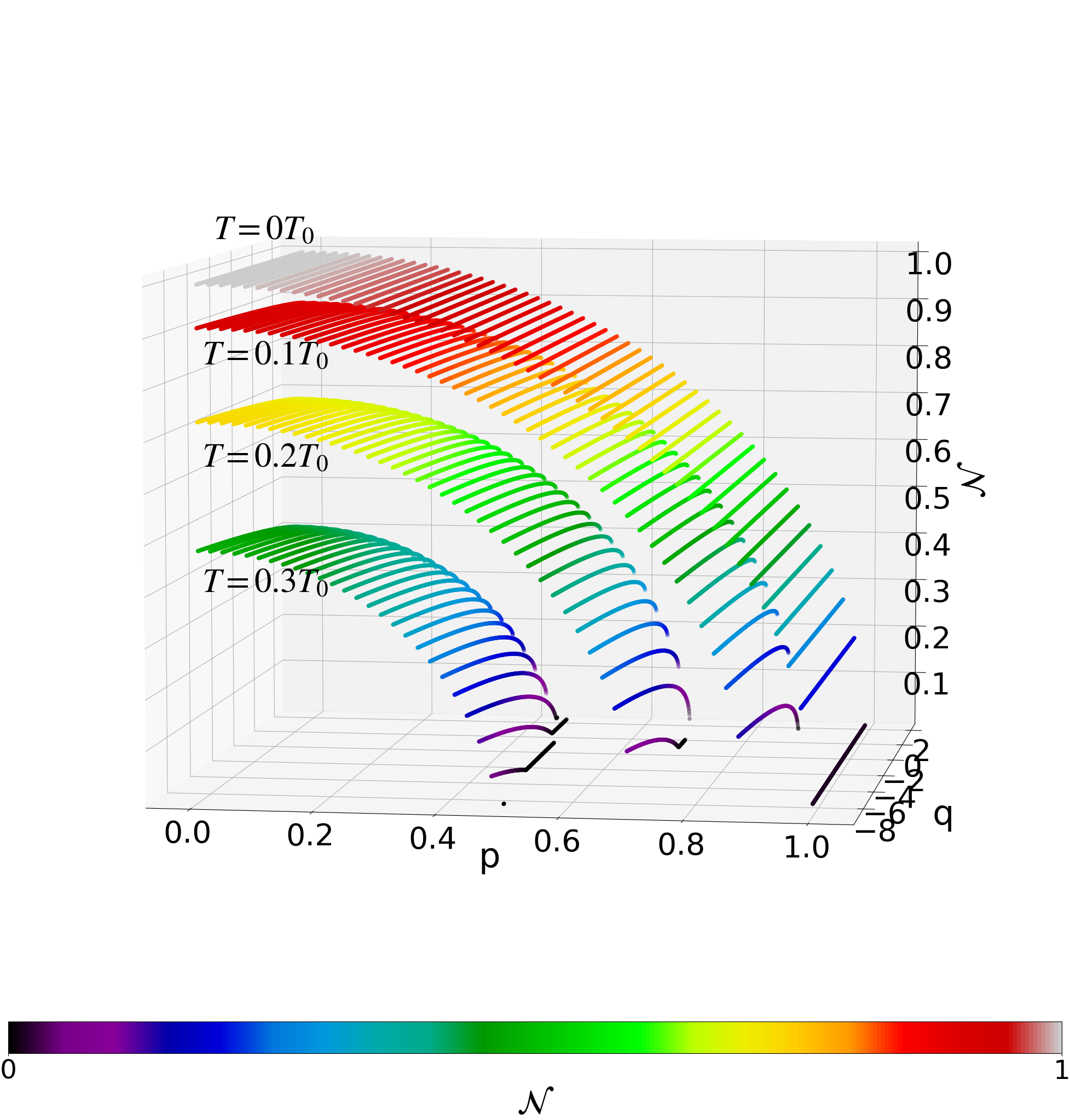}
    \caption{Negativity $\Nega$ of a BEC spin-1 on the polar phase at $T = 0,0.1, 0.2, 0.3$. At absolute zero (white surface), $\Nega=1$ and it is independent of the parameters $(q,p)$. At finite temperatures, on the other hand, the negativity decreases as one approaches to the regions: $p \to 1+q$, or $q<0$. }
    \label{Fig.2}
\end{figure}
%
%
%
%
\begin{figure}[t!]
    \centering
$p=0$
    
\includegraphics[width=.4\textwidth]{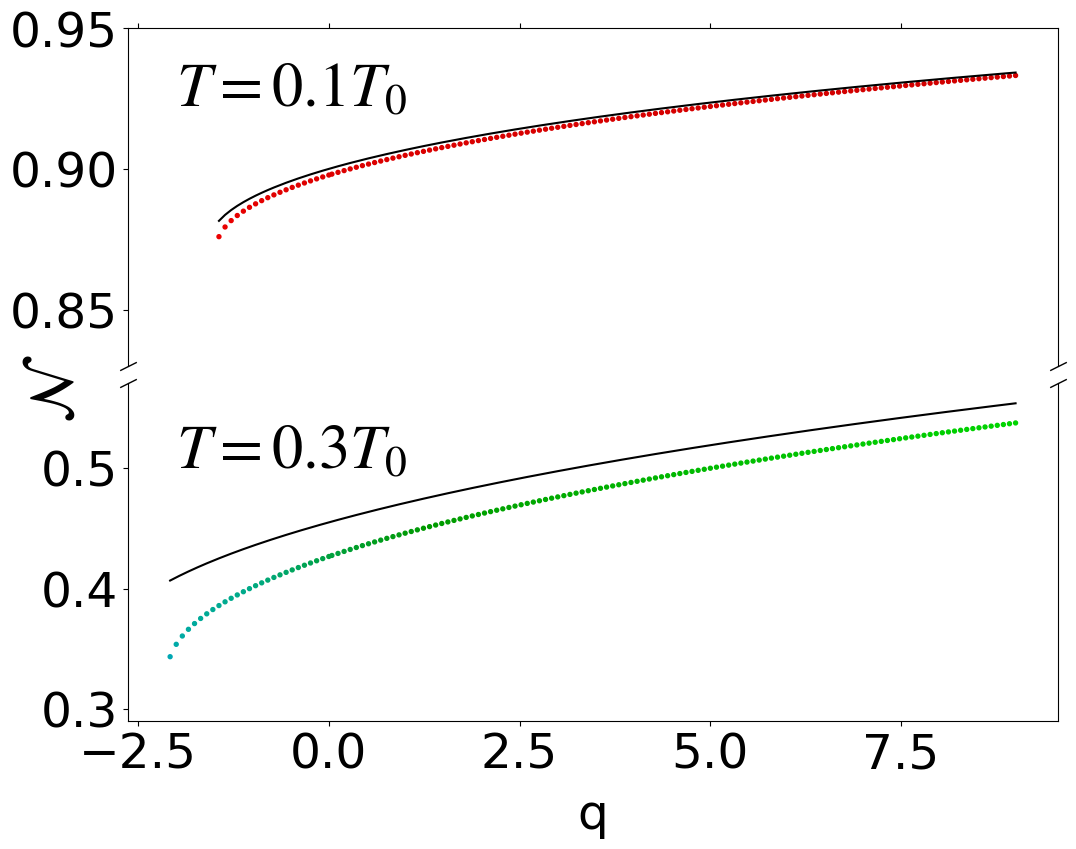}

$p=6$

\includegraphics[width=.4\textwidth]{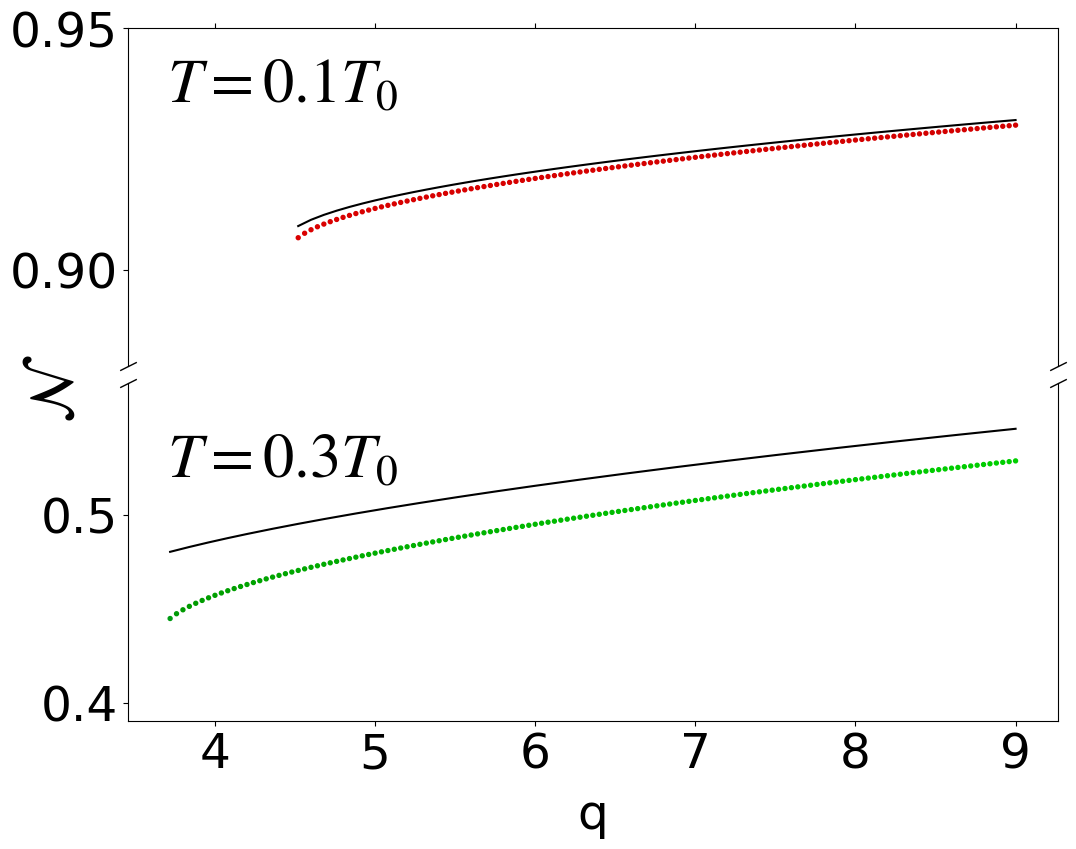}
    \caption{Negativity $\Nega$ of the polar phase as function of $q$ for $p=0$ (up) and $p=6$ (down) at $T=0.1$ and $T=0.3$. The numerical calculations (color dots) were done with the set of HF-GP equations, and the approximations (solid lines) are calculated via Eq.~\eqref{Neg.P.Ap}. }
    \label{Fig.3}
\end{figure}
%
%
In this section, we calculate numerically the entanglement $\Nega$ in the spinor part of the BEC by using the exact HF-GP equations~\eqref{sca.GPE.eq} for each spin phase in its admissible region at finite temperatures~\cite{Ser.Mir:21}. We plot the phase diagrams at different temperatures in Fig.~\ref{Fig.1}. Each phase has an extra metastability region beyond their phase transitions~\cite{Ser.Mir:21}. We also use in this section the approximated equations explained in Subsection~\ref{SubSec.HF} to calculate analytical approximations. We then compare the analytical and numerical results.
\subsection{FM phase}
Using Eqs.~\eqref{Neg.eq} and \eqref{Density.matrix}, the entanglement of the FM phase at finite temperatures is given by the expression
\begin{equation}
\label{Neg.Dicke}
\Nega(\rho) =  \max \left[ 0 , \, \sqrt{f_0^2+ \left(f_1 -f_{-1} \right)^2} - f_1 - f_{-1}
\right]
\, .
\end{equation}
which reduces to zero at $T=0$, where $f_1 =1$ and $f_0 = f_{-1} =0$. According to \eqref{Neg.Dicke}, the condensate has non-zero entanglement if the second term on the right hand side of the last expression is positive. This occurs only when
\begin{equation}
\label{HFNegFMInequality}
f_{0}^2 > 4 f_{1} f_{-1}  
,
\end{equation}
However, numerical calculations reveal that $\Nega(\rho) =0$ throughout the entire region considered. This observation aligns with the fact that an increase in temperature implies an increment of state mixedness in the condensate, which consequently does not generate entanglement by itself.
\subsection{P phase}
\begin{figure}[t!]7
    \centering
    \includegraphics[width=.5\textwidth]{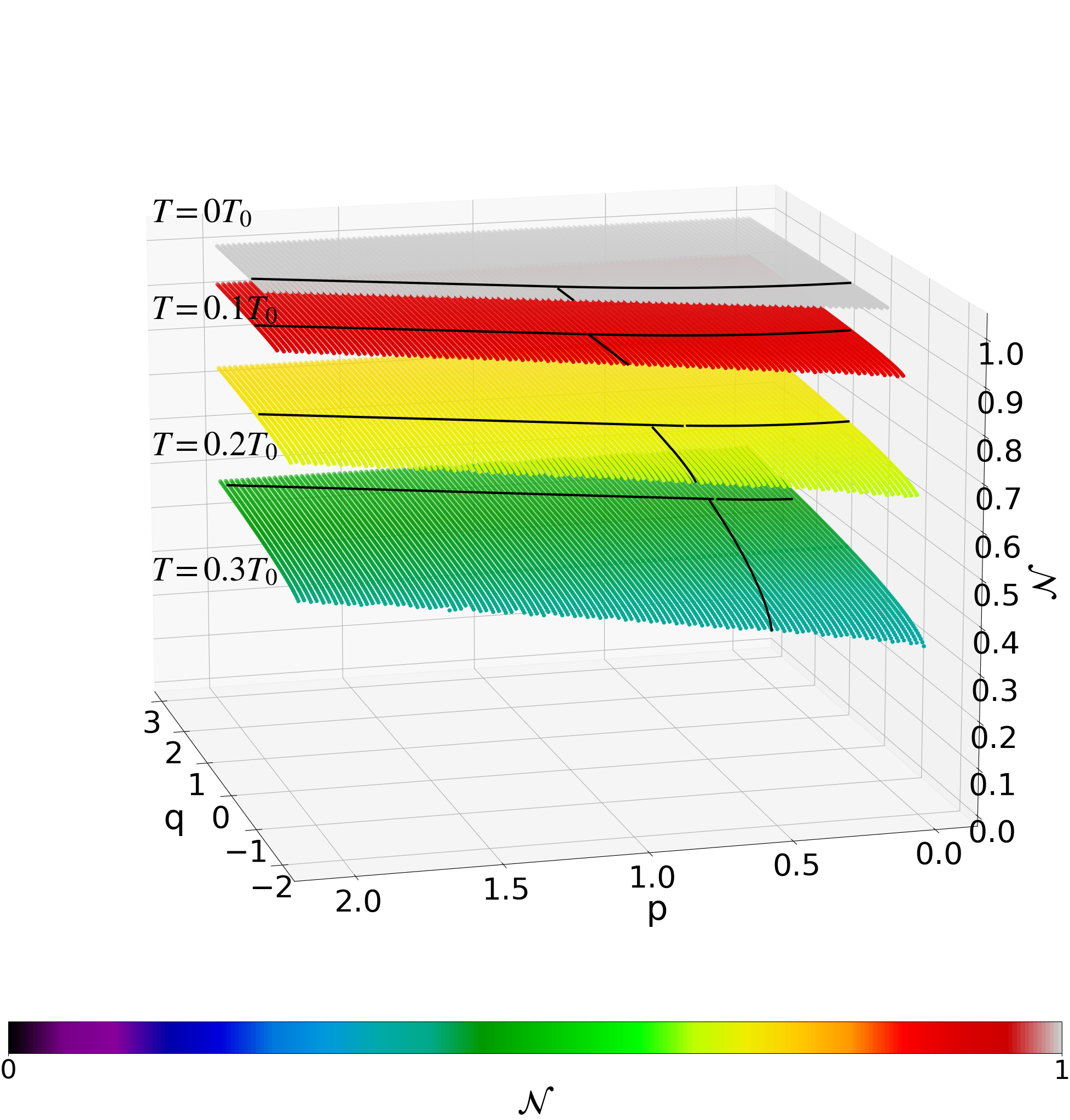}
    \caption{ Numerical calculations  of $\Nega$ of a spin-1 BEC on the AFM phase at $T = 0,0.1 , 0.2$ and $0.3$. At absolute zero, $\Nega$ follows the equation of a circle exactly with respect to $p$ for all $q$ values Eq.~\eqref{NegAFMMeanField}. At finite temperatures, the equation of a circle remains as a well approximation where the radius decreases as the temperatures increases. Additionally, it appears a fine dependency of $\Nega$ with respect to $q$. In general, the entanglement is inversely proportional to the parameters $(p,q)$.}
    \label{Fig.4}
\end{figure}
\begin{figure*}[t!]
\hspace{5.8cm} $T=0.1$ \hspace{4.2cm} $T=0.3$
\centering

\includegraphics[width=.95\textwidth]{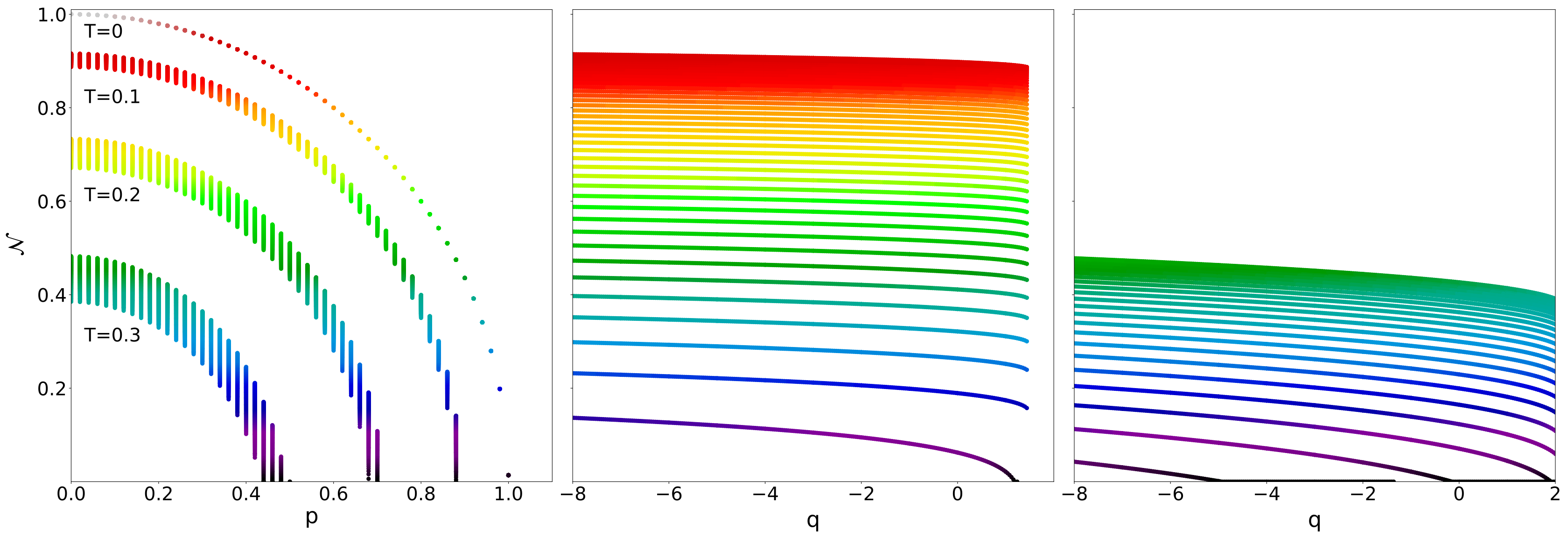}
    \caption{Projections of the Fig.~\ref{Fig.4} in the planes $p$ vs $\Nega$ (left) and $q$ vs $\Nega$ (center and right), respectively, at several temperatures.}
    \label{Fig.5}
\end{figure*}
We plot the entanglement as a function of the $(p,q)$ parameters of the P phase in Fig.~\ref{Fig.2}. The color density in the plot helps to visualize the value of $\Nega$ at different temperatures, specifically $T=0,0.1,0.2$ and $0.3$. Here, $\Nega$ is calculated numerically by using the HF-GP equations~\eqref{sca.GPE.eq}. We observe that, while the temperature significantly influences $\Nega$, its variation with respect to the $(p,q)$ parameters is comparatively smaller. Nonetheless, as temperature increases, the dependence of $\Nega$ on the $(p,q)$ parameters becomes more pronounced.

The approximated equations~\eqref{kappa.P} allow us to derive approximated analytical expressions for $\Nega$. For finite temperatures, $\Nega$ for the polar phase is equal to Eq.~\eqref{Neg.Dicke} with $f_0 = 1- f_1 - f_{-1}$ and $f_{\pm 1} = \La_{\pm 1}$. To scrutinize the entanglement in this spin phase, we take the second term on \eqref{Neg.Dicke} and rearrange it as follows:
\begin{equation}
    \left( \Nega + f_{1} + f_{-1} \right)^2 =  f_{0}^{2}+f_{1}^{2}+f_{-1}^{2} - 2f_{1}f_{-1} ,
\end{equation}
or using the relation $f_0 = 1- f_1 - f_{-1}$
\begin{equation}
    \left( \Nega +1\right)\left( \Nega-1\right) = -2\left( f_{1} + f_{-1} \right)\left( \Nega +1 \right) + \left( f_{1} - f_{-1}\right)^2 \, .
\end{equation}
In particular, since $\left(f_{1} - f_{-1}\right)^2 \ll 1$, the last expression is reduced to
\begin{equation}
\label{Neg.P.Supp}
    \Nega \cong 1 - 2\left(  f_{1} + f_{-1}\right) .
\end{equation}
By substitution of Eq.~\eqref{LambdaGenExp} on \eqref{Neg.P.Supp}, we obtained the reduced expression of $\Nega$ in terms of the $(q,p)$ parameters
\begin{equation}
    \label{Neg.P.Ap}
\begin{aligned}    
    \Nega = & 
    1 - 4k_2\Bigg[ T^{3/2} \zeta \left(\frac{3}{2}\right)
- k_1 \sqrt{T} \left( 1+q \right) \zeta \left(\frac{1}{2}\right)
\\
&    - \sqrt{\pi k_1 } T \left(\sqrt{1+q-p}+\sqrt{1+q+p}\right) \Bigg]  .
\end{aligned}
\end{equation}
We observe a good comparison between the numerical calculations and the analytical approximation~\eqref{Neg.P.Ap} in Fig.~\ref{Fig.3}.  Although the accuracy of the approximation diminishes as the temperature increases, the overall trend of $\Nega$ is captured with reasonable precision.
Since $\zeta \left(\frac{1}{2}\right)=-1.46035$ and $\zeta \left(\frac{3}{2}\right)=2.61238$, the last two terms proportional of Eq.~\eqref{Neg.P.Ap} may be negative.
These terms become negligible as one approaches to the regions i) $p\to1+q$, or ii) $(p,q) \rightarrow (0,-1)$, respectively, which are the phase transition zones of the P phase to the other two spin phases as shown in Fig.~\ref{Fig.2} (solid lines)~\cite{Ser.Mir:21}. To further explore the behaviour of $\Nega$, we compute its gradient in the space $(p,q)$, 
\begin{equation}
\begin{aligned}    
\label{GradNegPolar}
    \nabla \Nega = & \frac{\partial \Nega}{\partial p} \hat{p} + \frac{\partial \Nega}{\partial q}\hat{q}
\\
    = &2 k_2 \sqrt{\pi k_1} T \left( 
    \frac{\sqrt{1+q-p} - \sqrt{1+q+p}}{\sqrt{(1+q)^2-p^2}}
     \right) \hat{p}
\\
    &+ 2k_2 \Bigg[ \sqrt{\pi k_1}T\left(
    \frac{\sqrt{1+q-p} + \sqrt{1+q+p}}{\sqrt{(1+q)^2-p^2}}
    \right) 
\\
& + 2k_1\sqrt{T}\zeta\left(\frac{1}{2}\right) \Bigg]
    \hat{q} ,
\end{aligned}    
\end{equation}
revealing that the direction of maximum increase is given by $\nabla \Nega \propto -\hat{p} + \hat{q}$. A relative extremum, where $\nabla \Nega = \mathbf{0}$, occurs at $p=0$ and $q'=\frac{\pi T}{k_{1}}\frac{1}{\left| \zeta(\frac{1}{2}) \right|^{2}}-1$. At this point, 
\begin{equation}
\Nega_{\max}= 1 - \frac{4k_2 T^{3/2} \left[ \pi + \zeta\left( \frac{1}{2} \right) \zeta\left( \frac{3}{2} \right) \right]}{\zeta\left( \frac{1}{2} \right)}    .
\end{equation}
For example, at $T=0.1$ and $0.3$, $(q' , \Nega_{\max}) = (83.18 , 0.98)$ and $(251.53 , 0.88)$, respectively. The values obtained for $q'$ are quite overestimated with respect to those currently used in experimental settings~\footnote{According to the experimental values used in Ref.~\cite{stamper.Andrews.etal:1998}, the typical number of particles is $N=10^4$ with a density of $d=10^{20}m^{-3}$. This gives $c_1 N d /h = 547.7 \text{kHz}$. On the other hand, the typical values of the quadratic Zeeman fields $q/h \approx 8.9 \text{kHz}$~\cite{PhysRevLett.119.050404}. }. Nevertheless, we can conclude that, under laboratory conditions, the maximum entanglement in the polar phase at any finite temperature occurs for $p\to 0$ and $q \to \infty$. We can observe this trend in Figs.~\ref{Fig.2} and \ref{Fig.3}.
\subsection{AFM phase}
We plot the numerical results of $\Nega$ with respect to the $(p,q)$ parameters at various temperatures in Fig.~\ref{Fig.4}. Here, we use the HF-GP equations~\eqref{sca.GPE.eq}. Additionally, we plot its projections over the $\Nega$ vs $p$ and $\Nega$ vs $q$ in Fig.~\ref{Fig.5}. At $T=0$, we obtained an analytical expression $\Nega$ analytically in Eq.~\eqref{Eq.Nega.AFM.T0}, which can be reformulated as an equation of a circle with respect to $p$:
\begin{equation}
    \Nega^2 + p^2 = r^2 ,
    \label{NegAFMMeanField}
\end{equation}
with $r=1$, and without any dependency with respect to $q$. These properties are corroborated numerically in Figs.~\ref{Fig.4} and~\ref{Fig.5}. $\Nega$ has its maximum (minimum) when $p \to 0$ ($p \to 1$). At finite temperatures, the dependency of $\Nega$ with respect to $p$ remains well approximated by the equation of a circle with $r \approx 0.9, 0.7$ and $0.45$ for $T=0.1,0.2$ and $0.3$, respectively. It appears additionally a slight dependence on $q$ (see Fig.~\ref{Fig.5}). The radii values were calculated numerically by the least squares method. Similar to the P phase, the entanglement in the AFM phase decreases as $p$ increases. However, the opposite happens when we compare the dependency with respect to $q$ in both phases. For the AFM phase, $\Nega$ increases as $q$ decreases.
\section{Entanglement of the polar phase with respect to the spin-spin interactions}
\label{Sec.Ent.spin.spin}
Lastly, we investigate the analytical dependency of the nonlinear interactions in the HF-GP equations, given by the coupling factors $c_0$ and $c_1$, and the entanglement of a spinor BEC in the P phase. For this purpose, we reintroduce the bar notation to represent the scaled variables. We rewrite Eq.~\eqref{Neg.P.Ap} with the explicit dependence on the $c_k$ variables
\begin{equation}
    \label{Neg.P.Expanded}
\begin{aligned}    
    \bar{\Nega} = & 
    1 - \frac{4}{\lambda_0^3 N}\Bigg[ \bar{T}^{3/2} \zeta \left(\frac{3}{2}\right)
- \frac{\sqrt{\bar{T}}}{k_B T_0}  (c_1 N+q) \zeta \left(\frac{1}{2}\right)
\\
&    - \sqrt{\frac{\pi}{k_BT_0} } \bar{T} \left(\sqrt{c_1 N+q-p}+\sqrt{c_1N +q+p}\right) \Bigg]  .
\end{aligned}
\end{equation}
The last equation reveals two key observations: i) the entanglement within the spins of the condensate is independent of $c_0$ and ii) $c_1 N$ acts as an additional increment (shifting) of $q$. Consequently, the entanglement is expected to increase for BEC with stronger spin-spin interactions, indicated by a larger $c_1$. Furthermore, an increase in the total number of particles enhances the spin-spin interactions and consequently, also increases the entanglement.  
\section{Conclusions}
\label{Sec.Conclusions}
In this work, we have studied the spin entanglement of a spin-1 spinor BEC, defined via the formal
equivalence of the spin-1 states as two symmetric spin-1/2 states. We solve the Gross-Pitaevskii equations within the Hartre-Fock approximation for a spinor BEC of $^{23}$Na atomic species with spin-spin antiferromagnetic interactions, and study the thermal effects on the spin entanglement of the ground state of the condensate. By using the negativity of a two-qubit system as the measure of entanglement, we defined the notion of the spin entanglement and explored it in our BEC system and its dependence with respect to the temperature and the external Zeeman fields, $p$ and $q$, over its main magnetic phases: Ferromagnetic (FM), Polar (P) and Antiferromagnetic (AFM). Notably, the P and AFM phases present distinct features when temperature is considered. For the AFM phase, we obtain the entanglement and the linear Zeeman field are connected quadratically through the equation of a simple circle, in which its radius describes its effective temperature, and get reduced as the temperature increases. Conversely, the entanglement in the P phase behaves oppositely: it increases when $q$ increases. For the P phase, we also derive a simple analytical expression for the entanglement at finite temperatures, which fits with our numerical calculations. Additionally, we observed that the term $c_1 N$, where $c_1$ is the coupling factor associated with the spin-spin interactions, acts as an effective quadratic Zeeman field with respect to the entanglement. Thus, the spin entanglement increases as $c_1 N$ increases. 

The spin entanglement introduced in this work captures properties not detected by other kinds of entanglement, such as particle entanglement or mode entanglement. These findings could serve as a starting point to study the dependency of spin entanglement in spinor BECs and other many-body spinorial systems, with or without other complex interactions among their particles. As another important result of our work, we note that the spin-spin interactions in antiferromagnetic BEC within the P phase help to preserve the spin entanglement. The latter could have significant applications, for instance, in quantum metrology for the estimations of infinitesimal rotations. Future research might be extending this study by considering different interactions, other species of spinor BECs, or using other theoretical approaches beyond Hartree-Fock.
\section*{Acknowledgements}
ESE acknowledges support from the postdoctoral fellowship of the IPD-STEMA program of the University of Liège (Belgium). FM acknowledges funding from PAPIIT-DGAPA-UNAM through project number IN111624. 
\bibliographystyle{apsrev4-2}
\bibliography{refs_metastable}
\end{document}